# Suppression of threading defects formation during Sb-assisted metamorphic buffer growth in InAs/InGaAs/InP structure

A. Gocalinska[1,*], M Manganaro[1], and E. Pelucchi[1]

[1] *Tyndall National Institute, "Lee Maltings", University College Cork, Cork, Ireland*

**Abstract**

A virtual substrate for high quality InAs epitaxial layer has been attained via metalorganic vapor-phase epitaxy growth of Sb-assisted $In_xGa_{1-x}As$ metamorphic buffers, following a convex compositional continuous gradient of the In content from x = 53 % to 100 %. The use of trimethylantimony (or its decomposition products) as a surfactant has been found to crucially enable the control over the defect formation during the relaxation process. Moreover, an investigation of the wafer offcut-dependence of the defect formation and surface morphology has enabled the achievement of a reliably uniform growth on crystals with offcut towards the [111]B direction.

*Corresponding author

Low band-gap materials are of high interest, due to their applications i.e. in low-noise, low-power, high-frequency electron mobility transistors and infrared devices. The limited availability of suitable materials with large lattice parameters, which would serve as substrates, is a major draw-back. Epitaxy-ready wafers made out of InAs are expensive as such, and their limited dimensions are also elevating the costs of device production, as large-scale fabrication is impeded. A way of obtaining suitable materials for InAs lattice-matched overgrowth is to create a metamorphic buffer layer (MBL), in which the in-plane lattice parameter is being adjusted from more readily available material (like GaAs or InP). The objective is to obtain the highest surface quality (low density of defects in the final layer) with lowest possible cost, through reduction of the overall growth thickness.

Adjustment of the lattice parameter has been recently reported to give better morphological results, when the composition of the alloy follows exchange of group V, not III, elements[1], allowing also, by doing so, the control of the growth rate independent of the chemical composition (in a regime rate – limited by group III precursor dynamics). On the other hand $In_xGa_{1-x}As$ has historically been the material of choice, also because the control of (high) V/III ratio is often required for good growth quality. The competition between group V elements on the surface during the epitaxial growth in the first scheme discussed obviously makes this variable harder to control.

In relation to our contribution, due to a large lattice mismatch between InP and InAs (about 3 %), the $In_xGa_{1-x}As$ MBL bridging this gap suffers from high compressive strain. Relaxing the strain by defect formation allows for larger in-plane lattice parameters to be attained, but brings the well known risk of morphological disruptions due to formation of threading dislocations, which in turn give rise to high surface roughness. Control over the relaxation mode can be obtained in several ways, i.e. selecting specific gradient of compound change[2], allowing for self-curing of the system by thick overgrowth, reducing the growth temperature[3], and, in principle, any tool capable of modifying surface energies, like for example surfactant use, etc.

Alloy segregation in $In_xGa_{1-x}As$ is another possible source of surface roughening leading to formation of threading defects.[4] For that reason, compositionally graded MBLs are often found to be limited to low indium containing alloys. However, again, the use of surface acting agents may allow this obstacle to be overcome, as e.g. the surfactant effect of antimony was reported to be useful in controlling epitaxial growth mode, preventing 3-D growth in compressively strained InGaAs layers, mostly with applications to Quantum Wells (QW).[5]

In this letter we present important results on metalorganic vapor-phase epitaxy (MOVPE)-grown InGaAs metamorphic buffers with composition varying from lattice matched to indium phosphide (InP) up to indium arsenide (InAs). Extremely good surface quality and low residual strain was obtained after only 1.65 µm thick MBL growth. Strain release was controlled by a specific design of the gallium to indium exchange rate and the use of antimony as a surfactant proved to be an essential element.

Our findings represent a promising result towards the development of device grade metamorphic buffer layers, opening the way to a better and more effective exploitation of the 6.1 Å material family.

All epitaxial samples discussed here were grown in a high purity MOVPE[6] commercial horizontal reactor (AIX 200) at low pressure (80 mbar) with purified $N_2$ as carrier gas. The precursors were trimethylindium (TMIn), trimethylgallium (TMGa), trimethylantimony (TMSb), arsine ($AsH_3$) and phosphine ($PH_3$). The most successful sample design was as follows: 100 nm thick buffer layers of InP were grown on (001) InP perfectly oriented or slightly misoriented substrates (always with high tolerance, +/- 0.02 °) semi-insulating (iron doped) (the role of the substrate misorientation will become evident in the discussion in the last part of our manuscript); the buffer was followed by lattice-matched InGaAs 50 nm layer and 1650 nm TMSb assisted growth with indium concentration changing from 53% to 100%.

Growth conditions varied between the grown layers. The final optimised growth conditions were as follows (on 0.4 degrees misoriented substrates towards [111]B): InP buffers were grown with V/III ratio of 450 and growth rate of 1.4 μm/h at 700 °C thermocouple ($T_C$), according to our best established conditions for MOVPE grown InP.[7] InGaAs layers were grown with V/III of 130, at growth rate of 1 μm/h, with the temperature of 700 °C during the growth of lattice-matched buffer and dropping to 670 °C for the first 400 nm of graded growth, and then gradually reducing during the growth to 580 °C $T_C$ at the end of the grading. InAs cap layers, if present, were with 130 V/III ratio, 1 μm/h growth rate and at 580 °C $T_C$. The TMSb flow was switched on at the beginning of the graded layer at a value of $2\times10^{-5}$ mol/min (ratio As/TMSb ≈ 270) increasing during growth towards the end of the grading to the value of $4\times10^{-5}$ mol/min (ratio As/TMSb ≈ 135). The TMSb flow was not included into V/III estimations. Scanning transmission electron microscopy with energy-dispersive X-ray spectroscopic measurements (not shown) were performed after growth runs, showing no trace of Sb in the grown alloy, confirming the purely surfactant effect of the (TM)Sb. In structures capped with InAs layer, antimony support also was also used, with identical TMSb flow as at the end of the grading. All layers were nominally undoped.

Other growth efforts are also reported, with growth parameters described in detail when particular sample is discussed; in particular some reference samples were grown, without using TMSb as a surfactant during the graded layer growth.

All epitaxial growths resulted in smooth surfaces (see commentary on samples non-uniformities in the dedicated paragraph later in text) with cross-hatch pattern clearly visible when inspected with an optical microscope in (Nomarski) Differential Interference Contrast (N-DIC) or in dark field mode. Subsequent detailed morphological study was performed with Atomic Force Microscopy (AFM) in tapping/non contact mode at room temperature and in air. The defects formation and propagation was observed with cross sectional Transmission Electron Microscopy (TEM)

The assessment of composition and the strain in the layers was made according to measurements of Reciprocal Space Maps (RSM) obtained by high resolution X-ray diffraction measurements (HRXRD). Measurements were done in a symmetric (004) and two asymmetric (224 and -2-24) reflections with sample positioned at 0 °, 90 °, 180 ° and 270 ° with respect to its main crystallographic axes.

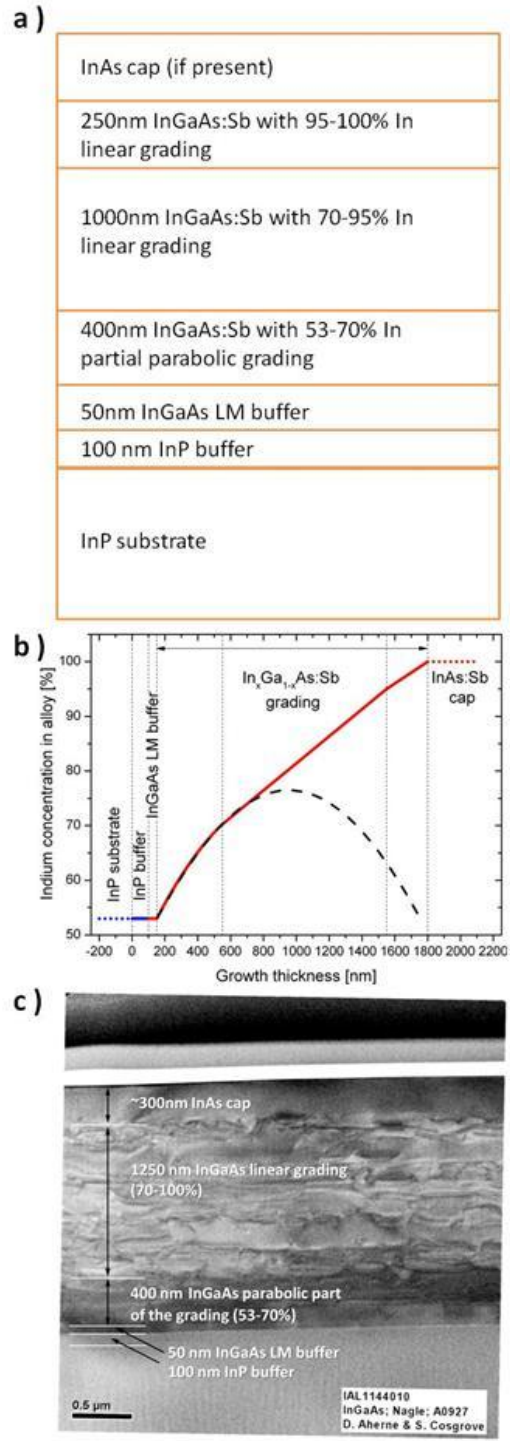

**Figure 1.** (Color online) (a) General sample structure. (b) The plot showing gradient of In-Ga exchange in the MBL (red trace); InP buffer marked with blue trace for indication only. Dashed curve show a full parabolic shape, part of which was selected as initial

MBL curve design. (c) TEM image in cross-section geometry of one of the realizations, showing defects distribution in MBL without their propagation into the cap.

On Figure 1 a) the sketch of our optimized sample design is presented. The choice of the exchange curve between Ga and In ions was made following past reported findings for the use of a parabolic grading as a convenient way for defect control.[8] We did observe by TEM imaging (not shown) a similar defect distribution as reported in Ref 8. On the steep part of the curve the defects formed rapidly, while when reaching the plateau the (mostly) non defected crystal structure was restored. However, if we kept the pure simple parabolic design as in Ref. 8 up to the InAs lattice parameter, the elastic strain, building up in the pseudomorphic part of the grading, caused a new generation of defects, threading to the sample surface and corrupting the final morphology. Furthermore, we also observed a very high residual strain which prevented reaching the in-plane InAs lattice parameter. Thus, to prolong the controlled defect formation range and improve the energy release, we modified the design and introduced the intermediate 1 μm thick part of linear grading, with slope as a tangent to prior parabolic curve, as sketched in Figure 1 b). This allowed for formation of intermediate density defects region, while the final 250 nm of the grading was grown on a linear, but slightly gentler grade, to reduce the dislocation density towards the final interface. (Figure 1 b)). When the MBL was overgrown with a cap layer of InAs, none of the previously present dislocations propagated into it, leaving the surface virtually defect-free, with normal step-bunched morphology (Figure 1 c)). It should be observed that we infer our surface defect density from TEM cross sectional analysis, and that this limits our detection to approx. $< 10^5$ defects/cm$^2$. Also, to grow this first sample discussed, we used a constant temperature of 700 °C throughout the buffers and grading and 600 °C for the cap growth, at variance to our optimised structure.

To qualitatively describe surface roughness we use the Root Mean Square (RMS) of the flattened AFM height scan in an area of 10×10 μm$^2$. The resulting surface morphology of the InAs overgrown layers presented an RMS of 13 nm, and a substantially step bunched surface (similarly to that observed in MOVPE low growth rate grown InP[7], also see Figure 2). Nevertheless the typical holes associated with threading dislocation (which would push the RMS to several tens of nm) appearing on the surface were missing, an indication that the surface organization is only linked to growth parameters and strain distribution, and not to a substantial presence of surface defects.

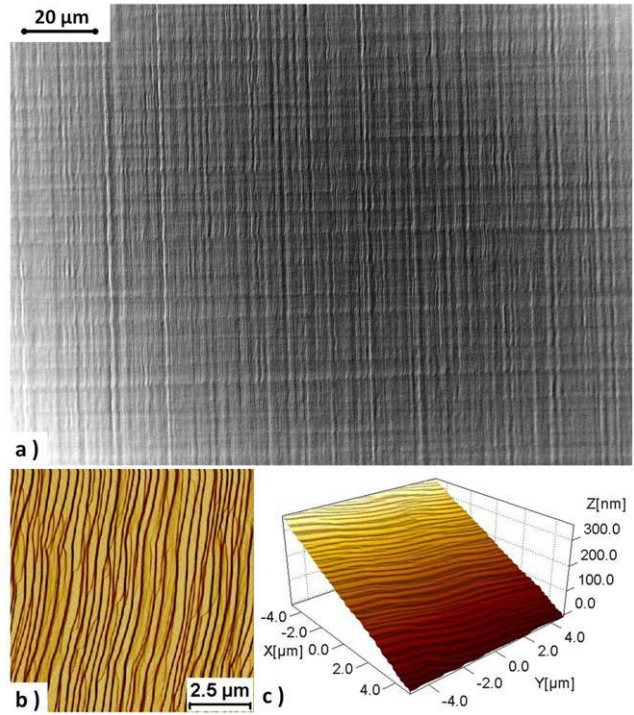

**Figure 2.** (Color online) Morphology of sample surface grown in optimized growth conditions on perfectly oriented substrate; (a) N-DIC image, AFM images: (b) signal amplitude and (c) 3D surface reconstruction.

Indeed, further optimization of the growth parameters allowed for growth of smoother surfaces with small residual strain: it was found that gradual reduction of the growth temperature led to less significant step-bunching and more defined step organization at the surface (Figure 2). Our optimized RMS is as low as 4 nm, which is mostly due to the ordered succession of step bunches, and not to defect related growth morphology. Basing on HRXRD RSMs, the parallel strain calculated[9] for cap layer overgrowing MBL was estimated to be $\varepsilon_p \approx -1.5\ \%$ ($\pm\ 0.5\ \%$ for different sample recipes, without substantial correlation to growth parameters), which corresponds to in-plane lattice parameter of 6.049 Å.

We should also say that during growth the MBL does not only relax the strain by defect formation, but, as observed in many other systems, also by tilting. The final crystallographic tilt observed by AFM on the samples was ~2° (vs. the [111]A direction as measured by HRXRD - see also later discussion and Figure 5).

To highlight the influence of Sb, we compared one of our optimized structures to an Sb free growth. Both of the samples discussed here were grown with 700 °C initial and 580 °C final growth temperature. The influence of Sb on the material quality was striking when the final morphology of the sample surfaces was compared (see Figure 3 c) and d), with a dramatic difference between the flow of step bunches, and a high density of defect related dimples). Moreover the in-situ monitoring of the sample performed during growth by reflectance at two different energies (see Figure 3 a) and b)) anticipated the differences between the two samples. While the 2.65 eV monitoring was sensitive to both surface quality and bandgap variation in the changing alloy, the 1.3 eV signal accounted mostly for the "mirror" quality of the wafer. Both traces show a significant drop of the recorded signal only in the Sb-free sample after about 1 μm of MBL growth, suggesting intensive defects formation from that point on. The signal recovered after another couple of hundred nanometers of growth, probably as a result of the relaxation processes in the structure, however the layer quality could not be restored even after the additional InAs cap overgrowth, an indication of a strong presence of surface threading defects. Note that the growth of the Sb-assisted sample was stopped immediately after MBL, without cap overgrowth, which would have planarised the surface even more − thus the difference in the behavior is even more striking.

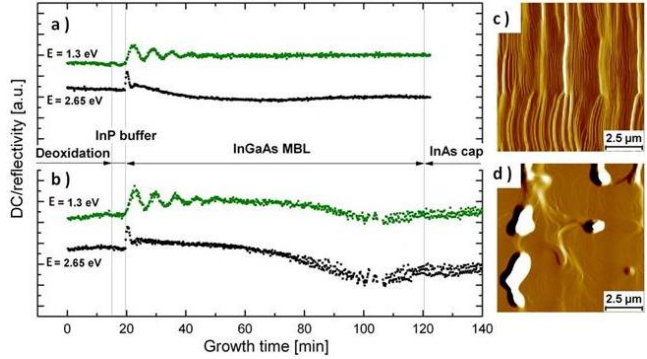

**Figure 3.** (Color online) (a) and (b) Reflectivity traces in respect to growth time for sample grown with (a) and without (b) Sb support. Morphology of respective surfaces (AFM signal amplitudes) are shown on (c) and (d). Sample grown with Sb (image c) had RMS of 7 nm, without Sb (d) RMS = 33 nm with holes depth exceeding 200nm, depth beyond standard AFM scanning abilities.

We want to stress a last point, which induced us to investigate lower growth temperatures, and different substrate miscuts. Despite the fact that the majority of the grown samples surfaces had good morphology, we have noted that in our initial growths conditions parts of the wafers were defected (the visual effect was evident directly with an optical microscope, see also Figure 4). In each case, the defected area of the sample had a shape of a roughly elongated "8" (with the vertical direction towards the [111]B planes), outside of which was the non-defected growth (pictorial sketch in Figure 4 a)). In the defected area the sample presented a high density of groove-like cracks along the [-110] direction, as shown in Figure 4 e)).

On close inspection, the step orientation of the surface layer was flowing radially from a single spot (Figure 4 b)), which after several tens of micrometers formed four distinct regions, with the steps rotated by 90 ° in respect to each other. When examined by means of HRXRD the crystallographic tilt was observed to be reversed in the defect-free parts which showed opposite step flow directions (the step flow mapped the tilt and vice-versa, and the two different tilts towards the opposite [111]A planes met at the centre spot, Figure 4 d)). Moreover, in the defected parts of the sample, the biggest tilt was off by 90 ° if compared to the good regions (i.e. towards the [111]B planes). What is worth noting, regardless of the size and type of used wafer piece, the "8" shape was constantly present and only appearing once on the sample, and thus its occurrence was clearly not correlated with any pre-growth wafer damage or impurity.

While we do not have a clear explanation on the origin of this peculiar singularity on our wafers, one possible cause of the observed problem could be strain induced wafer bending during the growth. This may well introduce locally different conditions and change the defects formation in the structure. Also temperature inhomogeneities across the wafer could be a possible result of different types of curvatures, apart from introducing strain related variation in the relaxation processes. Moreover, concave or convex deformation of the sample can lead to an increase or to a decrease of the local surface temperature of the wafer, leading to significantly different final result.[10]

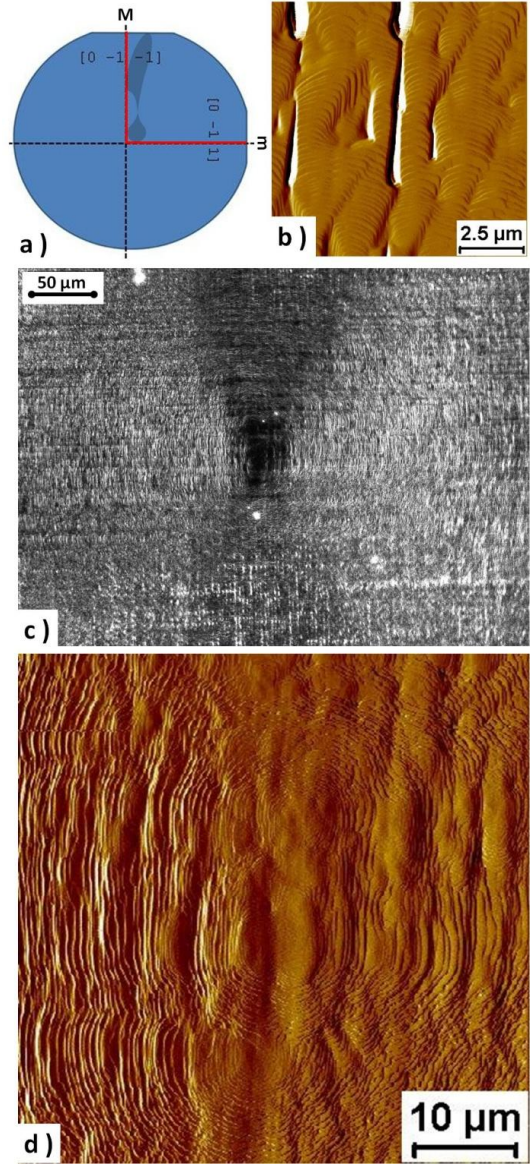

**Figure 4.** (Color online) (a) Sketch showing cleaved wafer piece used as a growth substrate, with defected part marked; M and m stand, respectively, for the major and minor flats of the 2″ wafer in EJ convention. The wafer was (100) with offcut 0.4 ° toward [111]B; (b) AFM image (signal amplitude) of defected parts of the wafer); (c) Dark field image of the defect center, (d) AFM image (signal amplitude) of the defect center.

To eliminate the singular defect formation, we varied our initial substrate miscut. Indeed, as the most of our growths were performed on quasi-singular wafers, it is possible that the real off-cut of the substrate varied across the surface, exposing A and B steps in different areas. To test this hypothesis, we performed a reference growth with higher off-cut wafers with intentional 0.4 ° miscuts toward [111]A and [111]B. We found that on purely [111]A misoriented substrates the growth was uniformly defected, reproducing the phenomenology of the "bad regions" obtained previously on perfectly oriented wafers (Figure 5 a)). Most of the [111]B misoriented growths presented low roughness (Figure 5 b)) (and the usual tilt towards the [111]A planes), although small defected patches could still be found on them (as in Figure 4). Only when the growth T was dropped as described in our experimental section, uniform, good quality surface of the whole [111]B misoriented wafer was obtained, eliminating the non-uniformity problem (when grown on quarter wafers, as we normally do). The crystallographic tilt measured by HRXRD on [111]B samples resembled closely that measured on good quality regions in perfectly oriented wafers (Figure 5 d) and f)), while the [111]A samples have shown almost no tilt, presumably due to the formation of defects which were efficiently releasing the accumulated strain energy (Figure 5 c) and e)).

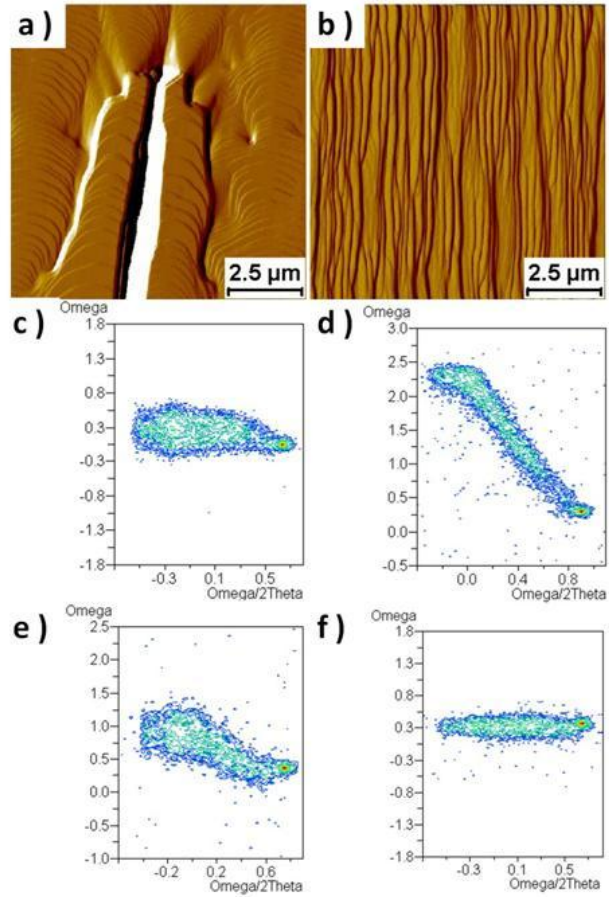

**Figure 5.** (Color online) Morphology and HRXRD RSMs of MBL grown simultaneously on (100) wafers with 0.4 ° toward [111]A (left panel) and [111]B (right panel) crystallographic directions. (a) and (b) show AFM images (signal amplitudes). RSMs (plotted in direct space) show the results of scanning in (004) reflection in 0 ° ((c) and (d)) and 90 ° ((e) and (f)) sample position in respect to beam incidence.

In conclusion we discovered that an appropriate combination of surfactant effects and original substrate miscuts are key ingredients to obtain high quality MBL. While more work will be needed to clarify the origin of the phenomenology described, the observed variations in the final results with relatively small substrate misorientation and temperature changes may justify the non-uniformity in literature reports on similar growth attempts as well as explain the generally poor state of the art reproducibility.

**Acknowledgments**

This research was enabled by the Irish Higher Education Authority Program for Research in Third Level Institutions (2007-2011) via the INSPIRE Programme, and partly by Science Foundation Ireland under grants 05/IN.1/I25, 10/IN.1/I3000 and 07/SRC/I117. We gratefully acknowledge Intel Corporation for the financial support and Intel Ireland for TEM analysis (Rob Dunne and Damian Aherne) and other support (Roger Nagle). The authors are grateful to Dr. V. Dimastrodonato for assistance with calculations, to Dr. P. Parbrook for useful discussion, and to Dr. K. Thomas for the MOVPE system support.